\documentclass[runningheads]{svmult}
\usepackage{makeidx}   
\usepackage{graphicx}  
\usepackage{subeqnar}  
\usepackage{multicol}  
\usepackage{cropmark} 
\usepackage{physprbb}  
\newcommand{\ltilde}
 {~ \raisebox{-1ex}{$\stackrel{\textstyle <}{\sim}$} ~}
\begin{document}
\title*{Dust versus Supernova Cosmology
}
\toctitle{
}
%
%
\titlerunning{}
%
\author{Tomonori Totani
}
\authorrunning{T. Totani}
%
%
\institute{
National Astronomical Observatory, Mitaka, Tokyo 181-8588, Japan
}

\maketitle              

\begin{abstract}
Here we present some critical discussions about the systematic 
uncertainty by dust extinction in the recent cosmological
results of high-redshift Type Ia supernovae. First we argue that
the currently available data do not robustly exclude the 
cosmologically significant extinction 
either by the reddening check or the dispersion argument
because of the observational uncertainties, even in the
case of ordinary dust that reddens. Then we discuss two theoretical
possibilities that high-$z$ supernovae have larger extinction 
and hence they are fainter than
local supernovae: the intergalactic dust and host galaxy
evolution. Optical and near-infrared observations for a large 
number of supernovae with improved
photometric accuracy are required to reject these possibilities
and derive a compelling cosmological result.
\end{abstract}

\section{Introduction}
Type Ia supernovae (SNe Ia) have been known as a representative
standard candle in the universe, and used in measurements of
cosmological parameters such as the Hubble constant ($H_0$), 
density parameter ($\Omega_M$), and the cosmological constant
($\Omega_\Lambda$). Recently two independent groups have obtained
the same result that a $\Lambda$-dominated flat universe is strongly
favored, by using a few tens of supernovae at redshift 
$z \sim$ 0.5 (Riess et al. 1998 [R98]; Perlmutter et al. 1999 [P99]).
Several possible sources of the systematic error in these measurements
have been discussed, including the Malmquist bias, $K$-correction,
supernova evolution, gravitational lensing, and so on.
The most serious among these is probably the effect of dust extinction.
Considering the very strong impact of the conclusion that our universe
is accelerating by vacuum energy density or dark energy, 
it is important to critically discuss how compelling these experimental 
results are, and discuss less exotic possibilities to explain 
the dimming of high-$z$ supernovae from various points of view.

Here we discuss the two points concerning the problem of dust extinction
in supernova cosmology: (1)
Do the current supernova data really exclude the extinction, 
either by ordinary dust or by speculative grey dust, affecting 
the cosmological result? (2) Are there theoretical expectations
that the extinction of high-$z$ supernovae is more significant than
that for local supernovae?  As an answer to the question (1),
we show that neither the color excess measurements nor the dispersion
argument robustly exclude the extinction that is comparable with the difference
between an open and $\Lambda$-dominated universes, even in the case of
ordinary dust that reddens. Then we discuss two theoretical
possibilities concerning the question (2), i.e., dust in intergalactic
field and evolution of
dust extinction by host galaxy evolution.

\section{Observational checks against the extinction effect: are they
compelling?}
\subsection{Reddening Check}
\label{section:reddening}
Both the two groups have made a considerable effort
to assess the systematic uncertainty due to extinction. For ordinary
dust that reddens, this effect can by checked by measuring the
difference of supernova colors at high-$z$ and local, provided that
the photometric measurement is sufficiently accurate. Both groups
reported that there is no significant color difference 
between the high-$z$ and local samples. However, it should be noted 
that there is considerable uncertainty in the color measurements
of high-$z$ as well as local supernovae, and an important question
is whether this uncertainty is sufficiently small compared with
the color difference induced by extinction that may affect the
cosmological results. The difference of an open universe and
$\Lambda$-dominated universe in the Hubble diagram is about 0.2 mag,
and extinction at the $B$ band with $A_B \sim$ 0.2 mag would
induce a color excess of $\Delta E(B-V) \sim$ 0.05 mag 
in $B-V$ color with the
standard extinction curve. Therefore supernova colors must be measured
with an accuracy much 
better than $\sim$ 0.05, for a compelling extinction check.

The mean $B-V$ colors of the high-$z$ R98 sample
is $-0.13\pm$0.05 or $-0.07\pm$0.05 depending on two analysis methods,
while expectation of unreddened color is $-0.10$ to $-0.05$.
Hence there is uncertainty of $\sim$ 0.05 mag in the color measurements
and this is significant for a cosmological test,
as shown above.  In the sample of P99, error-weighted
average reddening $\langle E(B-V) \rangle$ is 0.033$\pm$0.014 mag
for the local sample and 0.035$\pm$0.022 for the high-$z$ sample (P99).
This leads to the {\it statistical} uncertainty of at least $\sim$ 0.026
in the difference of extinction $\Delta E(B-V)$ between the high-$z$
and local sample. Therefore this test cannot exclude the cosmologically
significant extinction with confidence better than 2$\sigma$.
In addition to this, there should be {\it systematic} uncertainty in
$\Delta E(B-V)$; for example, P99 quote an overall systematic
uncertainty of 0.02 mag for the $K$-correction that may affect the
color measurements. The extinction observed in gravitational lens
galaxies also suggests that there may be systematic error in the
extinction estimate (see the last paragraph of this section).
Furthermore, it should be noted that this uncertainty
is reduced by statistics of a few tens of supernovae, and typical
error in the color measured for one supernova is typically
$\sim$ 0.4 mag in FWHM, that is much larger than the cosmologically important
color difference of 0.05 (see Fig. 6 of P99). If the probability distribution
of observational error is exactly the same for all supernovae,
the error in the estimate of mean color can be reduced by statistics,
but it is uncertain whether this condition is satisfied in the supernova
observations. 

P99 estimated the systematic uncertainty of extinction to be
less than 0.025 mag ($1\sigma$) in $A_B$, 
based on an analysis after removing nine reddest supernovae. 
Aguirre (1999b) argued that this limit does not apply if 
the dispersion in brightness and/or colors of high-$z$ supernovae
is dominated by factors other than extinction. 
As mentioned above, the typical observational 
uncertainty in $E(B-V)$ for one supernova is much larger than
the effect of cosmologically significant extinction.
Therefore, it is doubtful that P99 analysis successfully 
removed high-$z$ supernovae reddened by the systematic
effect of extinction. Rather, 
the supernovae removed by P99 might be reddened by 
intrinsic dispersion of supernovae, or simply by
observational errors.

In the analysis of R98, the reddening correction is systematically
included in the process of light-curve-shape fitting.
However, as noted above, the reddening induced by the cosmologically
significant extinction is comparable with or less than the observational error
of color measurement for one supernova.
In principle, it is difficult to correct the extinction effect
when the reddening is as small as the observational error in colors,
because the extinction correction is performed based on the observed colors.
The reddening correction by R98 will be effective for 
supernovae strongly reddened beyond the color uncertainty,
but it is not clear to us whether this correction has successfully corrected 
the systematic reddening that may affect the cosmological result.

Extinction observed in high-$z$ gravitational lens galaxies provides
us useful information about dust in host galaxies of supernovae.
Falco et al. (1999) compared their estimates
of extinction in lens galaxies in $0 \ltilde z \ltilde 1$ with
those estimated for the supernovae in the R98 sample. In spite of
comparable redshifts and impact parameters, they found 
the supernova sample to show markedly less extinction than the lens
sample. They suggested four possible interpretations of this
inconsistency. One of them is that the Type Ia fitting methods
are underestimating errors in extinction by a factor of 1.5--1.7.
Another is that an estimate of zero extinction in the supernova sample
may not really be zero. The Type Ia extinction estimates are actually
differential measurements relative to supernovae in nearby elliptical
galaxies that are defined to have zero extinction. However, 
lens galaxies, that are dominated by massive elliptical galaxies, 
do have some extinction. These results raise some doubts in the
extinction estimates of supernovae, and make the reddening check
by the two groups somewhat questionable.

Recently Riess et al. (2000) performed a near-infrared observation
and argued that this observation provides further evidence against
significant extinction in supernova cosmology.
We will discuss this point in \S \ref{section:discussion}.

\subsection{Dispersion Test}
The two groups (R98, P99) argued that the observed dispersion of apparent
magnitudes showing no significant evolution to high redshifts
gives further support that their results are not affected by
extinction. Their argument on the dispersion
test is valid if the observed dispersion is dominated by the
dispersion of extinction. R98 quoted a value of
dispersion $\sigma_{A_B} = 0.4$ mag when the mean extinction
is $\langle A_B \rangle = 0.25$ mag, based on a theoretical model of
spatial dust distribution by Hatano et al. (1998). If it is correct,
such a large dispersion is already in contradiction to what observed. 

However, there is large uncertainty
in dust distribution models. The detailed
procedure of deriving $\sigma_{A_B} = 0.4$ mag
is not clear in R98, but generally
the dispersion estimate is highly dependent on the theoretical modeling
as well as the selection effect. In fact, if the dispersion
$\sigma_{A_B} = 0.4$ is true, we should have already observed a
significant dispersion also in the local supernovae, because
various observations suggest that mean extinction in local galaxies
is typically about $\langle A_B \rangle \sim$ 0.1--0.2 mag
(see \S \ref{section:average_A_B}).
Since such a large dispersion is not observed in the local supernova
sample, it is suggested that the estimate of $\sigma_{A_B} = 0.4$
might be wrong.

Strongly extincted supernovae 
may not be observed below the detection limits, or may not be included
in the sample because they are clearly reddened. When such supernovae
are removed in the estimate of dispersion [i.e., like an
``extinction-limited subset'' in Hatano et al. (1998)], the 
expected dispersion becomes considerably smaller. For example, consider
the extinction distribution that is flat between a range 
$0 \leq A_B \leq 2 \langle A_B \rangle$. The dispersion of such a 
reasonable distribution is $\sigma_{A_B} = \langle A_B \rangle /
\sqrt{2}$, that is significantly smaller than the estimate of R98.
In this case, the dispersion becomes $\sigma_{A_B} = 0.14 
(\langle A_B \rangle / 0.2 {\rm mag})$ mag, that is comparable
with the intrinsic dispersion observed in the peak supernova magnitudes
after corrected by the light-curve-shape versus magnitude relation.
Therefore, it is possible that extinction is not a single dominant
origin in the observed dispersion and if it is the case the dispersion
test does not give a robust test of extinction.

In conclusion, it seems difficult to clearly check from the current data
a systematic change of average $B$-band extinction between the local
and high-$z$ sample that may affect the cosmological result, 
either by the reddening check or the dispersion test.
This conclusion applies to the ordinary dust that reddens with the
standard Galactic extinction curve, and
the check of possible greyer dust is, of course, even more difficult.

\section{Theoretical Possibilities for Larger Extinction of 
High-$z$ Supernovae}
Two theoretical possibilities have been proposed for the case
that high-$z$ supernovae have larger extinction than local supernovae:
the intergalactic dust (Aguirre 1999a, b) and host galaxy evolution
(Totani \& Kobayashi 1999). 

Aguirre (1999a, b) considered the effect of intergalactic dust
that may have been ejected from galaxies. It is known that a considerable
fraction of metals in galaxies in rich clusters is in a form of
metals in the intracluster medium. Renzini (1997) estimated that
metal mass in cluster gas is larger than that in member galaxies
by a factor of about 4. If the situation is the same for field galaxies,
a significant part of metals in the universe may be 
in the intergalactic field. 
If the dust mass in the intergalactic field in units of the critical
density is as large as $\Omega_{\rm dust} \sim 5 \times 10^{-5}$,
such intergalactic dust has an effect to change the cosmology
from a $\Lambda$-dominated flat universe 
into an open universe. The total metal mass
in the universe can be estimated by the cosmic star formation history,
and it is typically $\Omega_{\rm metal} \sim 2\times 10^{-4}$. Therefore,
if 25\% of metals in the universe is in the form of intergalactic
dust, it could have a significant effect on supernova cosmology.
Another important point of this idea is that such dust could 
have a significantly greyer extinction than the normal dust in galaxies.
If it is the case, it makes the reddening check even less powerful.
The dispersion argument discussed in the previous section 
does not apply to this possibility either, if the distribution of
intergalactic dust is sufficiently homogeneous.

This scenario is an interesting possibility, but it may be rather speculative.
It is highly uncertain whether the intracluster medium is
the same with the intergalactic field, and the mass fraction of dust
in all metals ejected from galaxies is also uncertain.
On the other hand, the other scenario of the host galaxy evolution
is less exotic and an effect that is certainly existent.
Even if supernovae themselves do not evolve to high redshifts, 
their host galaxies should undoubtedly evolve as shown by
various observations of galaxies at high redshifts. Both the
gas column density and gas metallicity, that are important physical
quantities for dust opacity, change with time by various star formation
histories depending on morphological types of galaxies.
Therefore, average dust extinction
in host galaxies must evolve systematically, and the question is
whether this effect is small enough not to affect supernova cosmology.
Unfortunately, it seems difficult to say the answer is yes,
from a viewpoint of the standard picture of galaxy evolution.

In the next section we describe our recent work (Totani
\& Kobayashi 1999), making
a quantitative estimate for this evolutionary effect by using
a realistic model of photometric and chemical evolution of galaxies
and supernova rate histories in various types of galaxies. We find that
typical evolution in average $A_B$ is $\sim$ 0.1--0.2 mag from $z=0$
to $\sim$0.5, 
that is significant for measurements of cosmological parameters
and may have escaped from the reddening check. Therefore, this effect 
should not be ignored in measurements of cosmological parameters
by high-$z$ supernovae.

\section{Evolution of Average Extinction in Host Galaxies}

Here we consider only the average extinction of a supernova in a
host galaxy, and do not consider the variation within a galaxy
depending on the supernova location in it.
Although the variation within a host galaxy can be washed out
by statistical averaging of many supernovae, evolution of galaxies
will cause systematic evolution of average extinction which cannot
be removed by statistical averaging. It is physically natural to 
assume that the dust-to-metal ratio is constant and hence 
the dust opacity is proportional to gas column density and 
gas metallicity of a host galaxy. In fact, it is well known that
the extinction in our Galaxy is well correlated to 
the HI gas column density (e.g., Burstein \& Heiles 1982; Pei 1992). 
It is also known that the dust opacity is correlated
to the metallicity among the Galaxy and the Large and Small
Magellanic Clouds, when gas column density is fixed (e.g., Pei 1992). 
Hence in the following we assume that the
dust opacity is proportional to gas column density and
gas metallicity, which evolve according to the star formation
history in a galaxy. 

The star formation history can be inferred from the present-day
properties of observed galaxies, by using the well-known technique of stellar
population synthesis. We can estimate the
time evolution of gas fraction and metallicity in a galaxy by using
photometric and chemical evolution models
for various galaxy types. Since the observed extinction of high-$z$
SNe Ia is an average over various types of galaxies,
we also need the evolution of SN Ia rate in various galaxy types.
In the next section, we describe the model of galaxy evolution and
SN Ia rate evolution used in this paper, which is constructed to
reproduce various properties of the present-day galaxies.

\subsection{Evolution of galaxies and Type Ia supernova rate}
We use photometric and chemical evolution models for five morphological types
of E/S0, S0a-Sa, Sab-Sb, Sbc-Sc, and Scd-Sd. The basic framework of the model
is the same as that of elliptical galaxies of Arimoto \& Yoshii
(1987) and that of spiral galaxies of Arimoto, Yoshii, \& Takahara (1992),
but model parameters are updated to match the latest observations
(Kobayashi et al. 1999), by
using an updated stellar population database of Kodama \& Arimoto (1997)
and nucleosynthesis yields of supernovae of Tsujimoto et al. (1995).
The model parameters for spiral galaxies are determined to reproduce 
the present-day gas fractions and $B-V$ colors in various galaxy types 
at 15 Gyr after the formation.
The model of elliptical galaxies is the so-called galactic wind model,
in which star formation stops at about 1 Gyr after the formation
by a supernova-driven galactic wind (Larson 1974; Arimoto \& Yoshii 1987). 
We assume that gas fraction in a elliptical galaxy decreases exponentially
after the galactic wind time ($\sim$ 1 Gyr), with a time scale same as
the galactic wind time. These models give the evolution of gas fraction
and metallicity in each galaxy type depending on the star formation
history.

SN Ia rate history in each type of galaxies is calculated with 
the metallicity-dependent SN Ia model introduced by Kobayashi et al. (1998).
In their SN Ia progenitor model, an accreting white dwarf (WD) 
blows a strong wind to reach the Chandrasekhar mass limit.  
If the iron abundance of progenitors is as low as [Fe/H]$\ltilde -1$, 
the wind is too weak for SNe Ia to occur. 
Their SN Ia scenario has two progenitor systems: one is a red-giant
(RG) companion with the initial mass of 
$M_{\rm RG,0} \sim 1 M_\odot$
and an orbital period of tens to hundreds days (Hachisu, Kato, \&
Nomoto 1996, 1999a).
The other is a near main-sequence (MS) companion with an initial mass of
$M_{\rm MS,0} \sim 2$--$3 M_\odot$
and a period of several tenths of a day to several days 
(Li \& van den Heuvel 1997; Hachisu et al. 1999b).
The occurrence of SNe Ia is determined from two factors: lifetime of
companions (i.e., mass of companions) and iron abundance of progenitors.
(See Kobayashi et al. 1998, 1999 for detail.)
This model successfully reproduces the observed chemical evolution in
the solar neighborhood such as the evolution of oxygen to iron ratio
and the abundance distribution function of disk stars (Kobayashi et al. 1998),
the present SN II and Ia rates in spirals and ellipticals, and
cosmic SN Ia rate at $z \sim 0.5$ (Kobayashi et al. 1999).

\subsection{Average extinction evolution towards high redshifts}
\label{section:average_A_B}
We have modeled the evolution of gas fraction ($f_{\rm g}$), 
metallicity ($Z$), and SN Ia rate per unit baryon mass of a galaxy
(${\cal R}_{Ia}$) in various types of galaxies, from which we calculate
the evolution of average extinction in the universe. 
We assume that these quantities do not depend on the mass of galaxies.
The basic assumption is that the dust opacity, and hence average
$A_B$ in a galaxy is proportional to gas column density and
gas metallicity. The average extinction
at redshift $z$ in a $i$-th type galaxy with the present-day $B$ luminosity
$L_B$ is given by $A_{B,i}(z, L_B) = \kappa f_{{\rm g},i}(t_z)
Z_i(t_z) [r_{{\rm e}, i}(L_B)]^{-2}  (M_{\rm b}/L_B)_i L_B$, where $t_z$ is 
the time from formation of galaxies,
$r_{\rm e}$ the effective radius of galaxies, and $(M_{\rm b}/L_B)$ is
the baryon-mass to light ratio which is determined by the evolution model.
We assume a single formation epoch $z_F$ for all galaxy types for 
simplicity.\footnote{In reality, there should be some dispersion in galaxy
ages. However, the systematic evolution of dust extinction 
is owing to the fact that all galaxies should become younger on average
towards high redshifts, and present-day 
age dispersion cannot remove this systematic 
effect.} The proportional constant $\kappa$ will be determined later.
We do not consider the size evolution of galaxies, and determine
$r_{\rm e}(L_B)$ from
empirical relations observed in local galaxies (Bender et al. 1992 for 
ellipticals, and Mao \& Mo 1998 for disk galaxies). 
It should be
noted that the extinction depends on the absolute luminosity of
galaxies. From the empirical $L_B$-$r_{\rm e}$ relation, the surface
brightness becomes brighter with increasing luminosity of disk galaxies,
and hence the massive galaxies should be more dusty than smaller ones.
This trend is consistent with observations (van den Bergh and Pierce
1990; Wang 1991).
Then the average extinction of SNe Ia over all galaxy types
at a given redshift is
\begin{equation}
\langle A_B(z) \rangle = \frac{ \sum_i \int dL_B 
A_{B, i}(z, L_B)
{\cal R}_{{\rm Ia}, i}(t_z) (M_{\rm b}/L_B)_i L_B \phi_i(L_B)}{
\sum_i \int dL_B {\cal R}_{{\rm Ia}, i}(t_z) (M_{\rm b}/L_B)_i L_B 
\phi_i(L_B)} \ ,
\end{equation}
where $\phi_i$ is the type-dependent galaxy luminosity function at
$z=0$, for which we adopted the Schechter parameters
derived by Efstathiou, Ellis, \& Peterson (1988)
using the catalog of the Center for Astrophysics (CfA) Redshift Survey
(Huchra et al. 1983).

We have to determine the overall normalization of extinction, $\kappa$, 
for which we use the average $V$ extinction of the Milky Way
(Sbc type, $L_B = 1.4 \times 10^{10}L_{B\odot}$) at $z=0$: $\langle
A_V \rangle_{\rm MW}$. This is an average of extinction of SNe Ia
occurring in our Galaxy seen by an extragalactic observer, and hence
it is different, in a strict sense, from the
average of the Galactic extinction which is extinction
of extragalactic objects observed by us. However,
if the location of the Sun is typical in the Milky
Way, we may infer this quantity by the average of the Galactic extinction.
The average Galactic extinction
of the 42 SNe Ia observed by P99 is $\sim$ 0.1 mag in $A_R$
or $A_I$ (see Table 1 of P99).
This suggests $\langle A_V \rangle_{\rm MW} \sim 0.1$--0.2 with the
standard Galactic extinction law (e.g., Pei 1992). 
The average reddening of the
Galaxy then becomes $\langle E(B-V) \rangle_{\rm MW} \sim$ 0.03-0.06 mag,
which is a typical reddening at the Galactic latitude of
$\sim$40--50$^\circ$ in the Galactic extinction map (Burstein \& Heiles 1982;
Schlegel, Finkbeiner, \& Davis 1998). This estimate is consistent
with a model of dust distribution in our Galaxy, which
suggests that the average extinction of SNe Ia 
in the Galaxy seen by an extragalactic observer
is typically $\langle A_V \rangle_{\rm MW} \sim$ 0.1--0.2 mag
(Hatano, Branch, \& Deaton 1998).\footnote{
Hatano et al. suggested that
most supernovae are only mildly obscured but
there is a long tail to stronger extinctions in the extinction
distribution. In the actual observations, 
such a tail will be cut out due to a magnitude limit of a survey. 
Hence, we have used here the mean extinction
of the ``extinction-limited subset'' in the Table 1 of Hatano et al., in which
strongly obscured supernovae with $A_B>0.6$ are removed.} 
Therefore we use $\langle A_V
\rangle_{\rm MW} \sim$ 0.1--0.2 mag as a plausible range of the
average extinction of our Galaxy.

Figure 1 shows the evolution of $B$ extinction for each galaxy type
as well as the average over all galaxy types, normalized by 
$\langle A_V \rangle_{\rm MW}$,
i.e., $\langle A_B(z) \rangle/\langle A_V \rangle_{\rm MW}$. 
Here we used a cosmological model with
$(h, \Omega_M, \Omega_\Lambda) = (0.5, 0.2, 0)$, and set $z_F = 4.5$
so that the age of galaxies is 15 Gyr which was assumed in
the evolution model. The thick solid line is
the average over all galaxy types, and the thin lines are for individual
galaxy types as indicated.
Since we have used a galactic wind model for elliptical galaxies,
they do not have interstellar gas and hence there is
no extinction in elliptical galaxies at $z<1$. 
(However, this may not be true as suggested by observations of
gravitational lens galaxies, see discussion in \S \ref{section:reddening}.)
The evolution of extinction is
caused by spiral galaxies, but the behavior of evolution is considerably
dependent on galaxy types. Early-type spiral galaxies become
more dusty towards $z \sim 1$, but an opposite trend is seen for late types. 
These behaviors can be understood as a competition of
the two effects: gas fraction evolution and metallicity evolution.
The gas fraction in early spiral galaxies is much smaller
than late types at present, but rapidly increases towards high
redshifts. This increase is responsible for increase of gas column
density and hence the dust opacity. On the other hand, the gas fraction
does not increase so much in late type galaxies, and decrease of metallicity 
towards high redshifts is responsible for the decrease
of dust opacity. In redshifts
more than 1, the extinction decreases towards higher redshifts 
in all spiral galaxies because the metallicity evolution becomes
dominant. 

\begin{figure}
\begin{center}
\includegraphics[width=.9\textwidth]{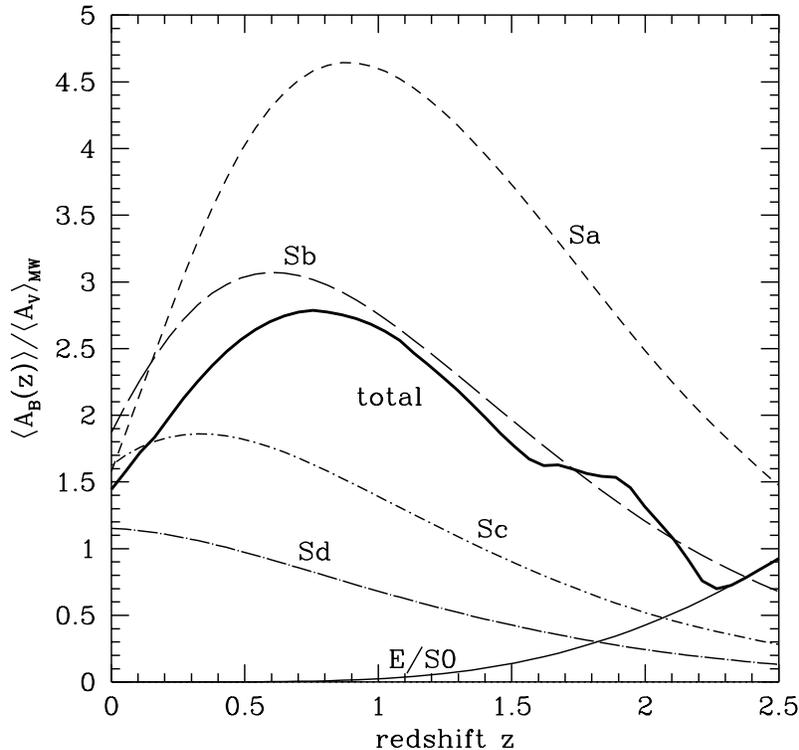}
\end{center}
\caption{Average $B$-band extinction of type Ia supernovae
as a function of redshift. Extinction $A_B$ is normalized by
the average $V$ extinction of our present-day
Galaxy, $\langle A_V \rangle_{\rm MW}$ 
(see text). The thick solid line is the average over the five morphological
types of galaxies, which is weighted by SN Ia rate in them.
Five thin lines are extinction evolution in individual galaxy types, as
indicated. An open universe with $(h, \Omega_M, \Omega_\Lambda) =
(0.5, 0.2, 0.0)$ is assumed, and the formation epoch of galaxies
is set to $z_F = 4.5$.}
\end{figure}

The average over all types is weighted by the SNe Ia rate
in each type. Because the star formation rate increases more rapidly
to high redshifts in early-type spiral galaxies than late types,
the average extinction is more weighted to early types at higher 
redshifts. Hence $\langle A_B \rangle / \langle A_V \rangle_{\rm MW}$ 
increases to high redshifts by $\sim 1$ from $z=0$ to 0.5.
This result suggests that, with $\langle A_V \rangle_{\rm MW} \sim$
0.1--0.2, the average extinction $\langle A_B \rangle$ 
of SNe Ia at $z \sim 0.5$ is larger than the local sample by about
0.1--0.2 mag. This systematic evolution of average extinction
is comparable with the difference between an open and
a $\Lambda$-dominated universe in the Hubble diagram, 
and hence this effect significantly
affects measurements of cosmological parameters. In the next section
we apply the above model in the estimate of cosmological parameters
by using the sample of P99.

\subsection{Effect on the Cosmological Parameters}
Figure 2 shows the Hubble diagram for SNe Ia of the
primary fit C of P99, which plots restframe 
$B$ magnitude residuals from a $\Lambda$-dominated flat 
cosmology [$(h, \Omega_M, \Omega_\Lambda) = (0.65, 0.2, 0.8)$] without
dust effect (thin solid line). 
Thin long- and short-dashed lines are the predictions
of the dust-free case with an open universe (0.5, 0.2, 0.0) and the
Einstein-de Sitter (EdS) universe (0.5, 1.0, 0.0), respectively. 
As reported by P99, the 
$\Lambda$-dominated flat universe gives the best-fit to the data.
Next, the thick lines show the predictions when the model of 
extinction evolution is taken into account, where
the cosmological parameters are the same with the dust-free curves
of the same line-markings. Here we adopt 
$\langle A_V \rangle_{\rm MW}$ = 0.2 mag.
In the open and $\Lambda$-dominated models, the galaxy formation epoch
is set to $z_F = 4.5$ and 5.0 so that the age of galaxies becomes
15 Gyr. In the EdS model, the age of the universe is shorter than 15 Gyr,
and hence we set $z_F = 5$ which gives an age of galaxies as
$\sim$ 12 Gyr. Although this age is a little shorter than that
assumed in the evolution model, the evolutionary effect
during 12--15 Gyr is small and hence this inconsistency is not serious.
As expected, the model curves with the dust effect are typically 
0.1--0.2 mag fainter than those without dust. As a result,
the open universe becomes the most favored cosmology among the three
when the extinction evolution is taken into account.

\begin{figure}
\begin{center}
\includegraphics[width=.9\textwidth]{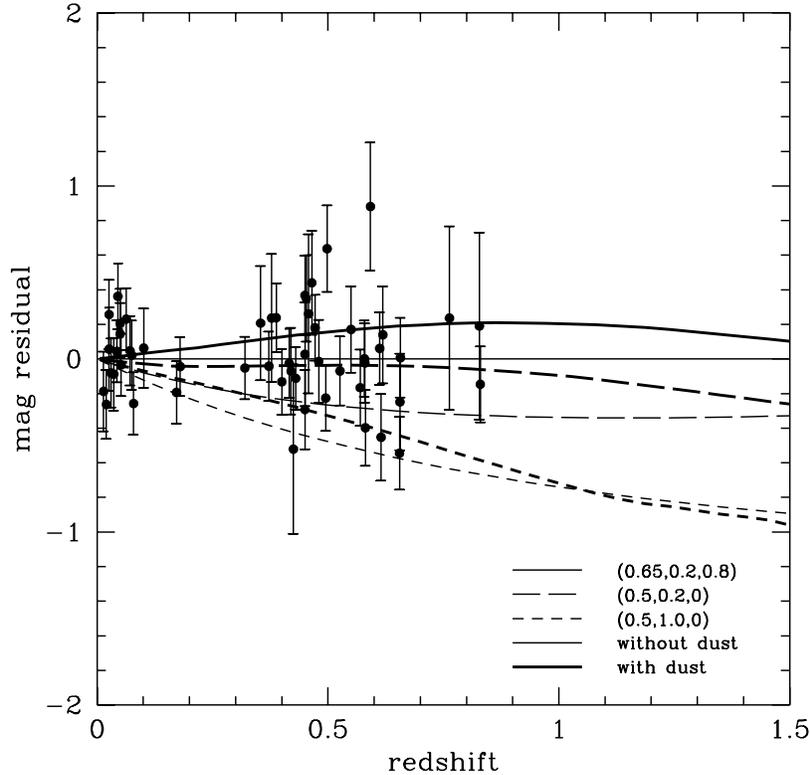}
\end{center}
\caption{The Hubble diagram of type Ia supernovae. The data are those
used in the primary fit C of Perlmutter et al. (1999). 
Restframe $B$ Magnitude residuals are from a $\Lambda$-dominated universe 
with $(h, \Omega_M, \Omega_\Lambda) = (0.65, 0.2, 0.8)$ without dust
effect (thin solid line). Thin long- and short-dashed lines 
are predictions
of an open universe (0.5, 0.2, 0.0) and Einstein-de Sitter universe
(0.5, 1.0, 0.0) without dust. The three thick lines are predictions
for the case with dust evolution, where the cosmological parameters
are the same with the dust-free curves of the same line-markings.
}
\end{figure}

We avoid more detailed statistical analysis to 
derive any decisive conclusion about the cosmological parameters,
because the result would be highly dependent on the
extinction evolution model. However, the evolution model presented
here is quite a natural and standard one without any exotic assumption.
Hence, our conclusion is that 
the systematic evolution of average extinction in host galaxies
should more carefully be taken into account when one uses 
SNe Ia to constrain the cosmological parameters.

\section{Discussion \& Conclusions}
\label{section:discussion}
Riess et al. (2000) reported near infrared observations 
of a supernova (SN1999Q) at $z = 0.46$. Such an observation
is useful to check the extinction, because the reddening of 
$(B-I)$ color is larger than $(B-V)$ color for the same amount of
dust and hence it is easier to detect. Riess et al. (2000)
found the reddening of $(B-I)$ color, i.e., $E(B-I) = -0.09\pm0.10$ 
mag for this supernova. This is consistent with almost no extinction
and inconsistent at the 3.4$\sigma$ confidence level 
with the reddening in this color of $E(B-I) \sim 0.25$ mag that is
required to save an open universe. However, it seems an overestimate 
that the reddening of $E(B-I) \sim$ 0.25 mag is required to save
an open universe. According to the
Galactic extinction curve of Savage \& Mathis (1979), it is found that
$A_B/E(B-I) = 1.58$. Then the $B$ extinction corresponding to 
$E(B-I)$ = 0.25 becomes $A_B$ = 0.39 mag. This $A_B$ seems larger
than the difference between an open and $\Lambda$-dominated flat
universe in the Hubble diagram, $\Delta m \sim 0.25$ mag 
at $z \sim 0.5$ (see Fig. 4 of R98). If we take $A_B
= 0.25$ as the extinction required to save an open universe,
corresponding reddening becomes $E(B-I)$ = 0.16 mag, and the confidence
level of inconsistency is lowered to 2.5$\sigma$. 

It still seems that there is 2.5 $\sigma$ inconsistency between
the observed infrared color and reddening to save an open universe, 
but it should be noted that 
any strong conclusion cannot be derived by only one supernova.
This argument may apply with only one supernova
to the intergalactic dust that is distributed homogeneously, 
but does not directly apply when there is scatter
in extinction of each supernova, as is the case of host galaxy
extinction. It is possible that SN1999Q had extinction
that is smaller than a mean value, due to its location in its host galaxy.
We must await a statistically large number of supernovae
whose colors are measured in infrared to conclude that the extinction
effect is negligible. The danger of discussing with a small number of
supernovae is obvious from the historical fact that
the supernova cosmology project originally claimed that the
$\Lambda$-dominated universe is disfavored with a smaller number of
supernovae (Perlmutter et al. 1997).

A more reliable way to constrain the cosmological parameters
by high-$z$ supernovae would be an analysis using only 
supernovae in elliptical galaxies, in which the dust evolution effect
is expected to be much smaller than spiral galaxies at $z<1$. 
(However, as discussed in \S \ref{section:reddening}, extinction observed in 
gravitational lens galaxies suggests that even such an analysis
may not be perfectly secure.)
In fact, P99 tried to analyze their supernovae
with known host-galaxy types, and found no significant change
in the best-fit cosmology. However, the host-galaxy classification
is only based on spectra of host galaxies without high-resolution
images. The uncertainty of a fit with a specified host-galaxy 
type is still large due to the limited number of supernovae, 
and P99 concluded that this test will need to await the host-galaxy
classification of the full set of high-$z$ supernovae and
a larger low-$z$ supernova sample. It is expected from our 
calculation that the fitting result of such an analysis in the future
will be dependent on host-galaxy types, giving important information
for chemical evolution of galaxies. High-$z$ supernovae beyond
$z \sim 1$ are also desirable to study galaxy evolution
as well as cosmological parameters.

We have argued that the observational check of dust extinction
effect on supernova cosmology, either by the reddening check or
dispersion argument, is not sufficiently compelling because of
the uncertainties in the photometric measurement of colors 
or large uncertainty of expected dispersions of extinction.
Observations of gravitational lens galaxies suggest that 
there may be some systematic uncertainties that have not yet been
identified in the extinction estimates of supernovae. 
On the other hand, it is theoretically not unreasonable that 
extinction of high-$z$ supernovae is larger than that of local
supernovae, by the intergalactic dust or host galaxy evolution.
These points should be kept in mind when one interprets 
the cosmological results
of high-$z$ supernovae. Optical and near-infrared observations for a large 
number of supernovae with improved
photometric accuracy are required to
derive a compelling cosmological result.


\begin{thebibliography}{8.}
\addcontentsline{toc}{section}{References}

\bibitem{a}
Aguirre, A.N., 1999a, ApJ, 512, L19

\bibitem{b}
Aguirre, A.N., 1999b, ApJ, 525, 583

\bibitem{ari87} 
Arimoto, N., \& Yoshii, Y. 1987, A\&A, 173, 23

\bibitem{ari92}
Arimoto, N., Yoshii, Y., \& Takahara, F., 1992, A\&A, 253, 21

\bibitem{bender}
Bender, R., Burstein, D. \& Faber, S.M. 1992, ApJ, 399, 462

\bibitem{c}
Burstein, D. \& Heiles, C. 1982, AJ, 87, 1165


\bibitem{c1}
Efstathiou, G., Ellis, R.S., \& Peterson, B.A. 1988, MNRAS, 232, 431

\bibitem{f99}
Falco, E.E. et al. 1999, ApJ, 523, 617

\bibitem{hkn99} 
Hachisu, I., Kato, M., \& Nomoto, K. 1999a, 
ApJ, 521, in press (astro-ph/9902304)

\bibitem{hknu99} 
Hachisu, I., Kato, M., Nomoto, K., \& Umeda, H. 1999b, 
ApJ, 519, in press (astro-ph/9902303)

\bibitem{d}
Hatano, K., Branch, D., \& Deaton, J. 1998, ApJ, 502, 177

\bibitem{d1}
Huchra, J.P., Davis, M., Latham, D. \& Tonry, J. 1983, ApJS, 52, 89

\bibitem{kob98} 
Kobayashi, C., Tsujimoto, T., Nomoto, K., Hachisu, I, \& Kato, M. 1998, 
ApJ, 503, L155

\bibitem{kob99} 
Kobayashi, C., Tsujimoto, T., \& Nomoto, K. 1999, to appear ApJ,
astro-ph/9908005

\bibitem{kod97} 
Kodama, T., \& Arimoto, N., 1997, A\&A, 320, 41

\bibitem{lar74} 
Larson, R. B., 1974, MNRAS, 169, 229

\bibitem{li97}
Li, X. -D., \& van den Heuvel, E. P. J. 1997, A\&A, 322, L9

\bibitem{mao}
Mao, S. \& Mo, H.J. 1998, MNRAS 296, 847 

\bibitem{e}
Pei, Y.C. 1992, ApJ, 395, 130

\bibitem{f}
Perlmutter, S. et al. 1997, ApJ, 483, 565

\bibitem{f1}
Perlmutter, S. et al. 1999, ApJ, 517, 565 (P99)

\bibitem{r97}
Renzini, A. 1997, ApJ, 488, 35

\bibitem{g}
Riess, A.G. et al. 1998, AJ, 116, 1009 (R98)

\bibitem{g1}
Riess, A.G. et al. 2000, ApJ, in press (astro-ph/0001384)

\bibitem{sm79}
Savage, B.D. \& Mathis, J.S. 1979, ARA\&A 17, 73

\bibitem{h}
Schlegel, D., Finkbeiner, D., \& Davis, M. 1998, ApJ, 500, 525

\bibitem{tk99}
Totani, T. \& Kobayashi, C. 1999, ApJ, 526, L65

\bibitem{tsu95}
Tsujimoto, T., Nomoto, K., Yoshii, Y., Hashimoto, M., Yanagida, S., \&
 Thielemann, F.-K. 1995, MNRAS, 277, 945

\bibitem{i}
van den Bergh, S. \& Pierce, M.J. 1990, ApJ, 364, 444

\bibitem{j}
Wang, B. 1991, ApJ, 383, L37

\end{thebibliography}
\end{document}